\documentclass[preprint,authoryear,12pt]{elsarticle}

\usepackage{graphicx}
\usepackage{epsfig}

\usepackage{amssymb}
\usepackage{natbib}
\usepackage{color}
\usepackage{tabularx}
\usepackage{float}

\newcommand       \mum        {\,{\rm \mu m}}

\newcommand       \Ks           {{\rm K_{S}}}

\newcommand       \simali       {\,{\sim}}
\newcommand       \magni        {\,{\rm mag}}

\newcommand       \amin       {a_{\rm min}}
\newcommand       \amax       {a_{\rm max}}
\newcommand       \Angstrom     {\,{\rm \AA}}

\newcommand \alphaS   {\alpha_{\rm S}}
\newcommand \alphaC   {\alpha_{\rm C}}
\newcommand \alphaX   {\alpha_{\rm X}}
\newcommand \acS        {a_{c,{\rm S}}}
\newcommand \acC        {a_{c,{\rm C}}}
\newcommand \acX        {a_{c,{\rm X}}}
\newcommand \g        {\,{\rm g}}
\newcommand \cm        {\,{\rm cm}}
\newcommand       \ppm          {\,{\rm ppm}}
\newcommand       \csun         {\left[{\rm C/H}\right]_{\odot}}
\newcommand       \cBstar       {\left[{\rm C/H}\right]_{\star}}

\newcommand       \fesun        {\left[{\rm Fe/H}\right]_{\odot}}
\newcommand       \feBstar       {\left[{\rm Fe/H}\right]_{\star}}

\newcommand       \sisun        {\left[{\rm Si/H}\right]_{\odot}}
\newcommand       \cism         {\left[{\rm C/H}\right]_{\rm ISM}}

\newcommand       \siism        {\left[{\rm Si/H}\right]_{\rm ISM}}
\newcommand       \cgas         {\left[{\rm C/H}\right]_{\rm gas}}

\newcommand       \cdust        {\left[{\rm C/H}\right]_{\rm dust}}
\newcommand       \sidust        {\left[{\rm Si/H}\right]_{\rm dust}}
\newcommand       \mgdust      {\left[{\rm Mg/H}\right]_{\rm dust}}
\newcommand       \fedust        {\left[{\rm Fe/H}\right]_{\rm dust}}
\newcommand       \xism        {\left[{\rm X/H}\right]_{\rm ISM}}

\newcommand       \xgas        {\left[{\rm X/H}\right]_{\rm gas}}
\newcommand       \xdust        {\left[{\rm X/H}\right]_{\rm dust}}
\newcommand \NS   {N_{\rm S}^\prime} % sil. dust column density;
\newcommand \NC   {N_{\rm C}^\prime} % carb. dust column density;
\journal{Planetary and Space Science}

\begin{document}

\begin{frontmatter}

\title{Modeling the Infrared Interstellar Extinction}
%\title{A plausible explanation of the flat mid-infrared
%       interstellar extinction}

\author[label1,label2]{Shu Wang}
\address[label1]{Department of Astronomy, Beijing Normal University,
                 Beijing 100875, China}

\author[label2]{Aigen Li}
\address[label2]{Department of Physics and Astronomy, University of Missouri,
                Columbia, MO 65211, USA}

\author[label1,label2]{B.W.~Jiang}

\begin{abstract}
How dust scatters and absorbs starlight in
the interstellar medium (ISM)
contains important clues about the size and composition of
interstellar dust. While the ultraviolet (UV) and visible interstellar
extinction is well studied and  can be closely fitted in terms of
various dust mixtures (e.g., the silicate-graphite mixture), the
infrared (IR) extinction is not well understood, particularly, the
mid-IR extinction in the 3--8$\mum$ wavelength range is rather flat
(or ``gray'') and is inconsistent with the standard
Mathis, Rumpl, \& Nordsieck (MRN) silicate-graphite dust model.
We attempt to reproduce the flat IR
extinction by exploring various dust sizes and species, including
amorphous silicate, graphite, amorphous carbon and iron.
We find that the flat IR extinction is best explained in terms of
micrometer-sized amorphous carbon dust
which consumes $\simali$60 carbon atoms
per million hydrogen atoms (i.e., C/H\,$\approx$\,60\,ppm).
To account for the observed UV/visible
and near-IR extinction, the silicate-graphite model requires
Si/H\,$\approx$\,34\,ppm and C/H\,$\approx$\,292\,ppm.
We conclude that the extinction from the UV to
the mid-IR could be closely reproduced by a mixture of
submicrometer-sized amorphous silicate dust,
submicrometer-sized graphitic dust,
and micrometer-sized amorphous carbon dust,
at the expense of excess C available in the ISM
(i.e., this model requires a solid-phase C abundance
of C/H\,$\approx$\,352\,ppm, considerably exceeding
what could be available in the ISM).
\end{abstract}


\end{frontmatter}

% \linenumbers

\section{Introduction\label{sec:intro}}
The interstellar extinction is one of the primary sources
of information about the interstellar dust size and composition.
The interstellar extinction varies from one sightline to another
in the ultraviolet (UV) and optical wavelength range. This variation
in the Milky Way galaxy can be described by a single parameter,
i.e., $R_{\rm V}$ (Cardelli et al.\ 1989; hereafter CCM).\footnote{%
   $R_V\,\equiv\,A_V/E(B-V)$ is the total-to-selective
   extinction ratio, where $E(B-V)\equiv A_B-A_V$ is the reddening
   which is the difference between the extinction
   in the blue band ($A_B$) and the extinction
   in the visual band ($A_V$).
   }
The average extinction law for the Galactic diffuse interstellar
medium (ISM) corresponds to $R_V\approx 3.1$.
Based on the interstellar extinction curve observed for
the diffuse ISM $R_V\approx 3.1$,
Mathis et al.\ (1977) constructed a simple interstellar dust model to fit
the interstellar extinction observed over the wavelength
range of $0.11\mum < \lambda < 1\mum$.
This classic model --- known as the ``MRN'' model --- consists
of silicate and graphite grains\footnote{%
   Hoyle \& Wickramasinghe (1962) first proposed that graphite
   grains of sizes a few times 0.01$\mum$ could condense
   in the atmospheres of cool N-type carbon stars,
   and these grains would subsequently be driven out of
   the stellar atmospheres and injected into
   the interstellar space by the stellar radiation pressure.
   Similarly, Kamijo (1963) suggested that nanometer-sized
   SiO$_2$ grains could condense in the atmospheres of
   cool M-type stars. Gilman (1969) argued that grains around
   oxygen-rich cool giants could mainly be silicates such as
   Al$_2$SiO$_5$ and Mg$_2$SiO$_4$.
   Silicates were first detected in emission
   in M stars (Woolf \& Ney 1969, Knacke et al.\ 1969).
   After blown out of the stellar atmospheres and
   injected into the interstellar space,
   silicates could become an interstellar dust component.
   Hoyle \& Wickramasinghe (1969) first modeled
   the interstellar extinction in terms of a mixture of
   silicate grains of radii $\simali$0.07$\mum$
   and graphite grains of radii $\simali$0.065$\mum$.
   Wickramasinghe \& Nandy (1970) found that a mixture
   of silicate, graphite, and iron grains achieved
   a rough fair fit to the interstellar extinction curve
   at $\lambda^{-1} < 8\mum^{-1}$.
   }
and takes a simple power-law size
distribution $dn/da \propto a^{-\alpha}$ with $\alpha\approx 3.5$
for the size range of $50\Angstrom < a < 0.25\mum$,
where $a$ is the radius of the dust
which is assumed to be spherical.\footnote{%
To be precise,
   the MRN model actually derived a {\it wider} size range
   of $50\Angstrom < a < 1\mum$ for the graphite component
   and a {\it narrower} size range of $0.025\mum < a < 0.25\mum$
   for the silicate component (and for other components such as
   SiC, iron and magnetite), with $\alpha$\,$\approx$\,3.3--3.6.
   In the literature, the MRN model is customarily taken
   to be a mixture of silicate and graphite with
   $\alpha=3.5$ and $50\Angstrom < a < 0.25\mum$.
   This is probably because (1) in their Figure~4
   the demonstrated model fit to the observed UV/visible
   extinction was provided by the olivine-graphite mixture
   with $\alpha=3.5$ and
   $50\Angstrom < a < 0.25\mum$ for both dust components;
   and (2) Draine \& Lee (1984) also derived
   $\alpha=3.5$ and
   $50\Angstrom < a < 0.25\mum$ for both dust components
   using improved optical constants for these two substances.
   The sudden cutoff at $\amin=50\Angstrom$ and
   $\amax=0.25\mum$ is not physical.
   Kim et al.\ (1994) and WD01 adopted a more smooth
   size distribution function which extends smoothly
   to $a>1\mum$. But the dust with $a>1\mum$ takes
   only a negligible fraction of the total dust mass.
   }
This model was further developed
by Draine \& Lee (1984) who extensively
discussed the optical properties of ``astronomical''
silicate and graphite materials.
Subsequent developments were made by
Draine and his coworkers
(Weingartner \& Draine 2001 [hereafter WD01], Li \& Draine 2001)
who extended the silicate-graphite grain model
to explicitly include polycyclic aromatic hydrocarbon (PAH)
molecules to explain the so-called ``unidentified
infrared emission'' (UIE) bands at 3.3, 6.2, 7.7, 8.6, and 11.3$\mum$
(see L\'eger \& Puget 1984, Allamandola et al.\ 1985).

With the wealth of available data from space-borne telescopes
(e.g., {\it Infrared Space Observatory} [ISO]
and {\it Spitzer Space Telescope})
and ground-based surveys (e.g., {\it Two Micron All Sky Survey} [2MASS])
in the near- and mid-infrared, in recent years we have seen an increase
in interest in the infrared (IR) extinction. Understanding the effects of
dust extinction in the IR wavelengths is important to properly interpret
these observations.
While the UV/optical extinction has been extensively
observed for a wide variety of environments and modeled in terms
of various dust models, our understanding of the near- and mid-IR
extinction is still somewhat poor and controversial,
despite that in this spectral domain many advances
have been made in the past few years
(see \S1 of Wang et al.\ 2013).

%%%% Figure 1 %%%%%
\begin{figure}[h!]
\centering
\vspace{-0.0in}
\includegraphics[angle=0,width=5.0in]{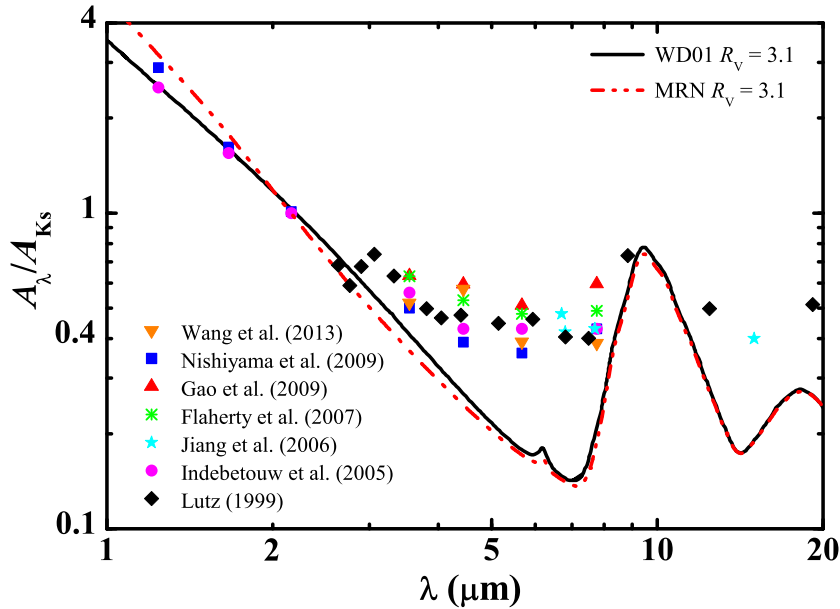}
\vspace{-0.1in}
\caption{\footnotesize
               \label{fig:IRExtObs}
               Comparison of the IR extinction observed for various
               interstellar regions with that predicted from the MRN
               (red dot-dashed line) and WD01 (black solid line)
               silicate-graphite models for the diffuse ISM of which
               the UV/optical extinction is characterized by
               $R_V\approx 3.1$.
               The little bump at 6.2$\mum$ arises from
               the C--C stretching absorption band of PAHs
               (see Li \& Draine 2001).
               }
\end{figure}

As shown in Figure~\ref{fig:IRExtObs}, WD01
silicate-graphite grain model predicts a power-law of
$A_\lambda \propto \lambda^{-1.74}$ for the IR extinction
at $1\mum < \lambda < 7\mum$, while the MRN model predicts
a steeper power-law of $A_\lambda \propto \lambda^{-2.02}$.\footnote{%
   At $\lambda>7\mum$, the extinction increases
   because of the 9.7$\mum$ silicate Si--O stretching
   absorption band.
   }
The model IR extinction curves reach their minimum at
$\simali$7$\mum$ where the extinction power-law intersects
the blue-wing of the 9.7$\mum$ silicate absorption band.

Rieke \& Lebofsky (1985) measured the IR extinction
from 1$\mum$ to 13$\mum$ for the lines of sight
toward $o$  Sco, a normal A5\,II star behind the edge
of the $\rho$ Oph cloud obscured by
$A_V\approx 2.92\magni$,\footnote{%
  The extinction toward $o$ Sco at $\lambda> 0.55\mum$
  can be well described by the $R_V=3.1$ extinction law.
  At $\lambda<0.55\mum$, the observed colors of $o$ Sco
  are much bluer than expected from those of a normal
  A5\,II star obscured by $A_V=2.92\magni$ with
  the $R_V=3.1$ extinction law,
  leading to the assignment of $R_V\approx 4.0$
  (see Rieke \& Lebofsky 1985).
  }
and toward a number of stars in the galactic center (GC).
Rieke \& Lebofsky (1985) derived a power-law of
$A_\lambda\propto \lambda^{-1.62}$ for
$1\mum < \lambda < 7\mum$ for
$o$ Sco and the GC sources.
Draine (1989) compiled the IR extinction observed
for a range of galactic regions including diffuse clouds,
molecular clouds, and HII regions. He derived
a power-law of $A_\lambda\propto \lambda^{-1.75}$ for
$1\mum < \lambda < 7\mum$.
More recently, Bertoldi et al.\ (1999) and Rosenthal et al.\ (2000)
also derived a power-law extinction of
$A_\lambda\propto \lambda^{-1.7}$
for $2\mum <\lambda < 7\mum$
for the Orion molecular cloud (OMC).\footnote{%
  The OMC extinction also displays an absorption band at 3.05$\mum$
  attributed to water ice.
  }

However, numerous recent observations suggest the mid-IR
extinction at $3\mum <\lambda< 8\mum$ to be almost
{\it universally} flat or ``gray'' for both diffuse and dense
environments (see \S1.4 of Wang et al.\ 2013 for a summary),
much flatter than that predicted from the MRN or WD01
silicate-graphite model for $R_V=3.1$
(see Figure~\ref{fig:IRExtObs}).

Lutz et al.\ (1996) derived the extinction toward the GC
star Sgr A$^{\ast}$ between 2.5$\mum$ and 9$\mum$
from the H recombination lines. They found that the GC
extinction shows a flattening of $A_\lambda$ in the wavelength
region of $3\mum <\lambda < 9\mum$,
clearly lacking the pronounced dip at $\simali$7$\mum$
predicted from the $R_V=3.1$ silicate-graphite model
(see Figure~\ref{fig:IRExtObs}).
This was later confirmed by Lutz (1999),
Nishiyama et al.\ (2009), and Fritz et al.\ (2011).

Indebetouw et al.\ (2005) used the photometric data
from the {\it 2MASS} survey and the {\it Spitzer}/GLIMPSE
Legacy program to determine the IR extinction.
From the color excesses of background stars,
they derived  the $\simali$1.25--8$\mum$ extinction laws
for two very different lines of sight in the Galactic plane:
the $l=42^{\rm o}$ sightline toward a relatively quiescent region,
and the $l=284^{\rm o}$ sightline which crosses the Carina Arm
and contains RCW~49, a massive star-forming region.
The extinction laws derived for these two distinct Galactic plane
fields are remarkably similar:
both show a flattening across the 3--8$\mum$ wavelength range,
consistent with that derived by Lutz et al.\ (1996) for the GC.

Jiang et al.\ (2006) derived the extinction at 7 and 15$\mum$
for more than 120 sightlines in the inner Galactic plane based
on the ISOGAL survey data and the near-IR data from DENIS and 2MASS,
using RGB tip stars or early AGB stars (which have only moderate mass
loss) as the extinction tracers. They found the extinction  well
exceeding that predicted from the MRN or WD01 $R_V=3.1$ model.

Flaherty et al.\ (2007) obtained the mid-IR
extinction laws in the {\it Spitzer}/IRAC bands
for five nearby star-forming regions.
The derived extinction laws
at $\simali$4--8$\mum$ are flat, even flatter than
that of Indebetouw et al.\ (2005).

Gao et al.\ (2009) used the {\it 2MASS} and {\it Spitzer}/GLIPMSE data
to derive the extinction in the four IRAC bands for 131 GLIPMSE fields
along the Galactic plane within $|l|\leq65^{\rm o}$.
Using red giants and red clump giants as tracers,
they also found the mean extinction
in the IRAC bands to be flat.

Wang et al.\ (2013) determined the mid-IR extinction
in the four {\it Spitzer}/IRAC bands of five individual regions
in Coalsack, a nearby starless dark cloud,
spanning a wide variety of interstellar environments
from diffuse and translucent to dense clouds.
They found that all regions exhibit a flat mid-IR extinction.

All these observations appear to suggest
an ``universally'' flat extinction law in the mid-IR,
with little dependence on environments.\footnote{%
  We should note that an ``universally'' flat
  mid-IR extinction law does not necessarily mean
  an identical mid-IR extinction law for all regions,
  instead, it merely means a flattening trend of
  $A_\lambda$ with $\lambda$ in the mid-IR.
  Chapman et al.\ (2009), McClure (2009),
  and Cambr\'{e}sy et al.\ (2011) found that
  the shape of the mid-IR extinction law appears
  to vary with the total dust extinction.
  But also see Rom{\'a}n-Z{\'u}{\~n}iga et al.\ (2007)
  and Ascenso et al.\ (2013) who found no evidence
  for the dependence of the mid-IR extinction law
  on the total dust extinction.
  }
While rapid progress has been made in observationally
determining the mid-IR extinction and numerous IR
extinction curves have been accumulated both for
the Milky Way and for the Magellanic Clouds
(e.g., see Gao et al.\ 2013a), theoretical understanding
of the nature and origin of the flat mid-IR extinction
lags well behind the observations. We plan to model
the interstellar extinction from the far-UV to the far-IR
for a wide range of interstellar environments.
In this work we present our first attempts to understand
the nature of the flat mid-IR extinction.
In \S\ref{sec:status} we summarize the current status in
interpreting the flat mid-IR extinction.
\S\ref{sec:mod} describes our model.
\S\ref{sec:rst} presents the results and discussions.

%%%% Figure 2 %%%%%
\begin{figure}[h!]
\centering
\vspace{-0.0in}
\includegraphics[angle=0,width=5.0in]{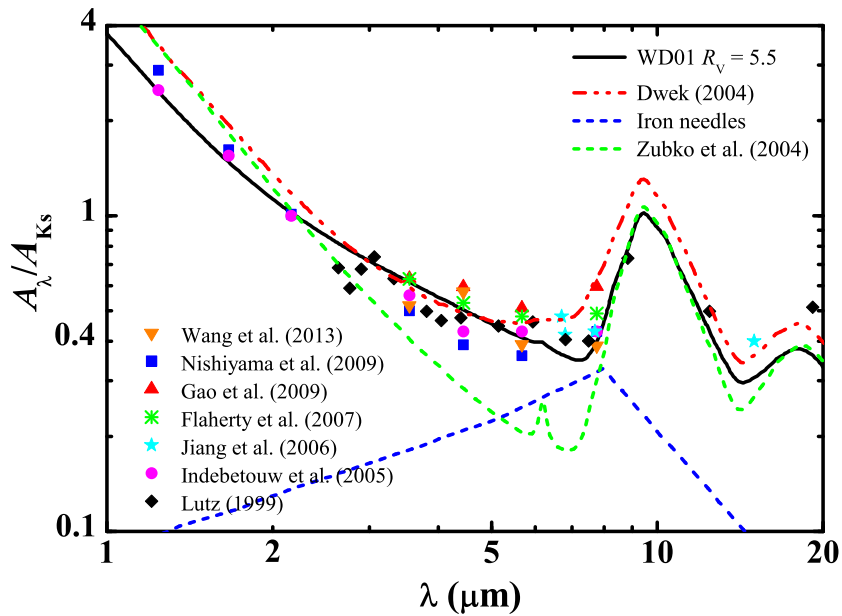}
\vspace{-0.1in}
\caption{\footnotesize
               \label{fig:IRExtMod}
               Comparison of the IR extinction observed for various
               interstellar regions with that predicted from the WD01
               model for $R_V=5.5$ (black solid line) and the iron
               needle model (red dot-dashed line) of Dwek (2004)
               which is a combination of the $R_V=3.1$
               silicate-graphite model of Zubko et al.\ (2004; green
               dashed line) and iron needles (blue dashed line).
               }
\end{figure}

\section{Modeling the Mid-IR Extinction:
            Where Do We Stand?\label{sec:status}}
To the best of our knowledge, only WD01 and Dwek (2004)
have closely reproduced the flat mid-IR extinction.
Using a mixture of amorphous silicate dust and
carbonaceous dust,\footnote{%
   The carbonaceous grain population was assumed to
   extend from grains with graphitic properties at radii
   $a>0.01\mum$, down to particles with PAH-like
   properties at very small sizes (see Li \& Draine 2001).
   }
WD01 fitted the UV/optical and near-IR extinction curves
of different values of $R_V$ in the wavelength range of
$0.125\mum < \lambda < 2.86\mum$.
Draine (2003) extended the WD01 model of $R_V=5.5$
into the IR up to $\lambda < 30\mum$.
It is amazing that the WD01 $R_V=5.5$ model closely
reproduces the flat mid-IR extinction observed toward the GC
(Lutz et al.\ 1996; also see Figure~\ref{fig:IRExtMod}).
The success of the WD01 $R_V=5.5$ model
and the failure of the WD01 $R_V=3.1$ model in fitting the flat
mid-IR extinction suggest that the flat mid-IR extinction implies
a population of large dust, although the exact size and quantity
of this dust population are not known.

Generally speaking, denser regions tend to have flatter extinction
curves in the UV and higher values $R_V$:
increased $R_V$ is attributed to tilt in size distribution to
decrease numbers of small grains, and increase of numbers
of large grains due to grain growth primarily through coagulation
(Draine 2011).

However, the flat mid-IR extinction is not only seen in dense regions.
As discussed in \S\ref{sec:intro}, the flat mid-IR extinction has been
seen in various interstellar environments, including diffuse clouds
(e.g., the low-density lines of sight in the Galactic midplane
[see Zasowski et al.\ 2009], and the diffuse and translucent regions
of the Coalsack nebula [see Wang et al.\ 2013]).
The $R_V=5.5$ size distribution is more appropriate for dense regions.

Dwek (2004) hypothesized that metallic needles may play an important role
in accounting for the flat mid-IR extinction.
As shown in Figure~\ref{fig:IRExtMod}, Dwek (2004) argued that,
combined with the silicate-graphite mixture for the $R_V=3.1$ model
(Zubko et al.\ 2004), metallic needles\footnote{%
    The idea of metallic needles was originally brought up
    by Hoyle et al.\ (1968) and Wickramasinghe et al.\ (1975)
    to explain the 2.7\,K cosmic microwave background (CMB).
    They argued that the 2.7\,K CMB might have arisen from
    the radiation of ``Population III'' objects thermalized
    by long slender conducting cosmic whiskers or ``cosmic needles''.
    }
with a typical length ($l$)
over radius ($a$) ratio of $l/a\approx600$
and a needle-to-H mass ratio of $\sim5\times10^{-6}$
could explain the flat mid-IR extinction derived by Lutz (1999) for
the GC. However, it is not clear if metallic needles are indeed
present in the ISM (see Li 2003). It would be interesting to study
their generation and evolution in the ISM and their optical properties.

Gao et al.\ (2013b) have also attempted to fit
the $\simali$1--19$\mum$ IR extinction curve
toward the GC derived by Fritz et al.\ (2011).
Their best-fit model for the GC IR extinction constrains
the visual extinction to be $A_V$$\simali$38--42$\magni$.
But their model could not simultaneously reproduce both
the relatively {\sl steep} $\simali$1--3$\mum$ near-IR extinction
and the {\sl flat} $\simali$3--8$\mum$ mid-IR extinction.
They suggested that the extinction toward the GC could be
due to a combination of dust in different environments:
dust in diffuse regions (characterized by small $R_V$
and steep near-IR extinction),
and dust in dense regions (characterized
by large $R_V$ and flat UV extinction).

\section{Our Model\label{sec:mod}}
We aim at simultaneously reproducing the observed UV/optical and near-
and mid-IR extinction. Following MRN and WD01, we assume a mixture of
amorphous silicate dust and graphite dust.
The optical constants of amorphous silicate dust
and graphitic dust are taken from Draine \& Lee (1984).
Since the flat mid-IR extinction is seen
both in diffuse clouds and in dense clouds,
we consider the UV/optical and near-IR extinction curves
of different $R_V$ values:  $R_V = 3.1, 4.0$, and 5.5.

We adopt an exponentially-cutoff power-law size distribution
for both dust components:
$dn/da \propto a^{-\alpha}\exp\left(-a/a_c\right)$
for $a_{\rm min } < a < a_{\rm max}$,
where $a$ is the spherical radius of the dust
(we assume the dust to be spherical),
$\alpha$ is the power exponent,
and $a_c$ is the cutoff size.
Following MRN, the lower cutoff of the dust size
is initially taken to be $a_{\rm min}=50\Angstrom$.
But for the $R_V=4.0, 5.5$ curves, we find that
$a_{\rm min}=25\Angstrom$ is more favorable.
The upper cutoff is set at $a_{\rm max}=2.5\mum$.

We have five parameters: $\alpha_{\rm S}$ and $a_{c,{\rm S}}$
for the silicate component,
$\alpha_{\rm C}$ and $a_{c,{\rm C}}$ for the carbonaceous
component (i.e., graphite),
and $f_{\rm C2S}$, the mass ratio of graphite to silicate dust.
The ratio of the total model extinction at wavelength $\lambda$
to the column density of H nuclei is calculated from
\begin{equation}
A_{\lambda}/N_{\rm H} = 1.086\,\NS \times C_{\rm ext}({\lambda}) ~~,
\end{equation}

\begin{eqnarray}
\nonumber
C_{\rm ext}(\lambda) & = &
\int_{\amin}^{\amax} C_{\rm ext,S}(a,\lambda)
a^{-\alphaS}\exp\left(-a/\acS\right)\,da\\
& + & \left(\NC/\NS\right)
\int_{\amin}^{\amax} C_{\rm ext,C}(a,\lambda)
a^{-\alphaC} \exp\left(-a/\acC\right)\,da ~~,
\end{eqnarray}
\begin{eqnarray}
\nonumber
\NS &=& \sum\limits_{j=1}^{N_{\rm obs}}
\left\{(A_{\lambda}/N_{\rm H})_{{\rm obs},j} \times
C_{\rm ext}(\lambda_j)/\sigma_{{\rm obs},j}^2\right\}\\
&& /\left[1.086\,\sum\limits_{j=1}^{N_{\rm obs}}
\left\{C_{\rm ext}(\lambda_j)
/\sigma_{{\rm obs},j}\right\}^2\right] ~~,
\end{eqnarray}
\begin{eqnarray}
\nonumber
\NC/\NS &=& f_{\rm C2S} \times
\left(\rho_{\rm S}/\rho_{\rm C}\right) \\
& & \times
\int_{\amin}^{\amax} a^{3-\alphaS} \exp\left(-a/\acS\right)\,da
/\int_{\amin}^{\amax} a^{3-\alphaC}
\exp\left(-a/\acC\right)\,da ~~,
\end{eqnarray}
where $\NS$ and $\NC$ are respectively
proportional to the column densities
of the silicate and graphite dust,
$\rho_{\rm S}=3.5\g\cm^{-3}$ and $\rho_{\rm C}=2.24\g\cm^{-3}$
are respectively the mass densities
of the silicate and graphite dust,
$C_{\rm ext,S}(a,\lambda)$ and $C_{\rm ext,C}(a,\lambda)$
are respectively the extinction cross sections of
the silicate and graphite dust of
size $a$ at wavelength $\lambda$,
$C_{\rm ext}(\lambda)$ is the extinction
at wavelength $\lambda$ of
the silicate-graphite mixture per silicate dust column,
and $\left(A_{\lambda}/N_{\rm H}\right)_{{\rm obs},j}$ and
$\sigma_{{\rm obs},j}$ are respectively the ``observed'' extinction
and uncertainty at wavelength $\lambda_j$.
For the UV/optical and near-IR extinction,
we follow WD01 to fit the CCM extinction characterized by $R_V$
between 0.35 and 8$\mum^{-1}$
and evaluate the extinction at $N_{\rm obs}=100$ wavelengths
$\lambda_j$, equally spaced in $\ln\lambda$.
%%%

As will be shown later in \S\ref{sec:rst}, although this simple
grain model could closely reproduce the observed UV/optical
and near-IR extinction, it could not fit the flat mid-IR extinction.
Guided by the fact that dust scatters and absorbs starlight
most effectively at wavelengths comparable to its size
(i.e., $2\pi\,a/\lambda \sim1$; see Wang et al.\ 2014),
we will add an extra population of large, micrometer-sized dust
to account for the flat mid-IR extinction.
For this large, $\mu$m-sized dust population,
we will consider amorphous silicate, graphite,
amorphous carbon, and metallic iron.
We adopt the optical constants of the ``ACAR''-type
amorphous carbon of Rouleau \& Martin (1991).
The optical constants of iron are taken from
Li (2003).

For the mid-IR extinction, we compare the model results
with the observational data for the following regions
as a whole:
the Galactic center (Lutz 1999, Nishiyama et al.\ 2009),
the Galactic plane (Indebetouw et al.\ 2005,
Jiang et al.\ 2006, Gao et al.\ 2009),
nearby star-forming regions (Flaherty et al.\ 2007),
and the Coalsack nebula (Wang et al.\ 2013).
Ideally, the CCM UV/optical extinction should be combined
only with the IR extinction for the regions characterized
by the same $R_V$. However, the regions for which the mid-IR
extinction has been measured are not well characterized
and their $R_V$ values are not well determined.\footnote{%
   For example, the line of sight toward the Galactic center
   is often considered as a sightline of diffuse clouds
   as revealed by the presence of the 3.4$\mum$ C--H aliphatic
   hydrocarbon stretching absorption feature which is not seen
   in molecular clouds
   (Pendleton \& Allamandola 2002, Mennella et al.\ 2002).
   However, the Galactic center sightline must also contain
   molecular cloud materials as revealed by the presence of
   the 3.1$\mum$ and 6.0$\mum$ H$_2$O ice absorption features
   (e.g., see McFadzean et al.\ 1989).
   }
Nonetheless, as illustrated in Figure~\ref{fig:IRExtObs},
the mid-IR extinction for different regions does not
appear to differ substantially from each other
and on average, a flat mid-IR extinction is well established.
It is interesting to note that,
as shown in Wang et al.\ (2013) for the Coalsack nebula,
the mid-IR extinction of diffuse regions
seems to be even somewhat flatter than that of dense regions.
Therefore, we intend to compare our model results
with the mid-IR observations as a whole,
instead of an individual region.
%%%

In modeling the UV/optical extinction,
we perform a grid-search by minimizing $\chi^2$
(see eq.\,8 of WD01) with the weights given by WD01.
We fit the mid-IR extinction by eye
as we do not know how the mid-IR extinction data points
shoud be weighted compared to the UV/optical extinction.
We fit the UV/optical segment and the mid-IR segment
separately but requiring that the addition of
the mid-IR model extinction to the total extinction
does not distort the fit to the UV/optical part.

%%%% Figure 3 %%%%%
\begin{figure}[h!]
\centering
\vspace{-0.0in}
\includegraphics[angle=0,width=5.0in]{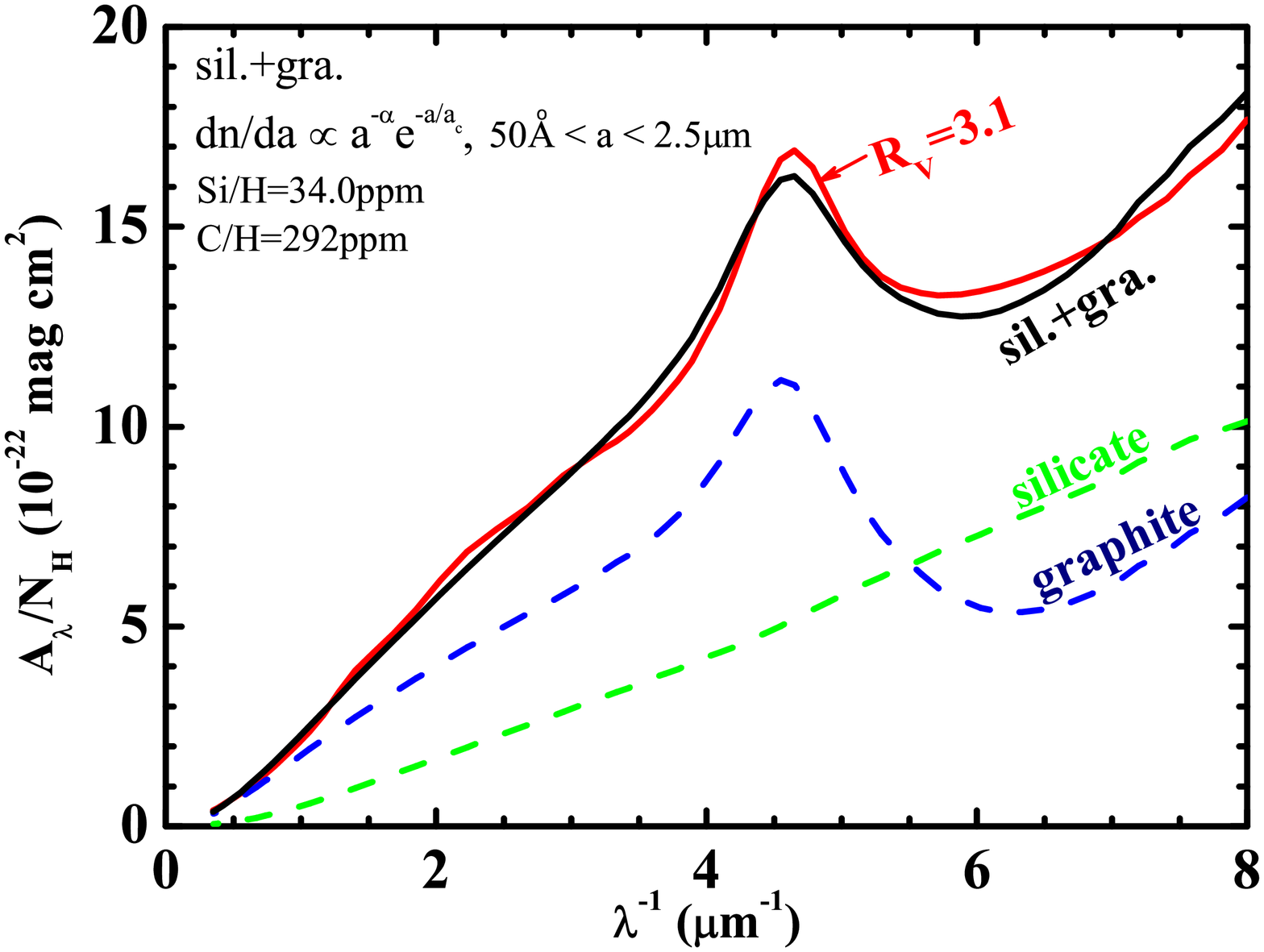}
\vspace{-0.1in}
\caption{\footnotesize
               \label{fig:WLJUVOpt}
               Fitting the $R_V=3.1$ UV/optical and near-IR extinction
               with a simple mixture (black solid line)
               of amorphous silicate (green dashed line)
               and graphite dust (blue dashed line).
               The black solid line plots the model fit,
               while the red solid line plots the $R_V=3.1$
               extinction curve observed for diffuse regions.
               See \S\ref{sec:mod} for details.
               }
\end{figure}

%%%% Figure 4 %%%%%
\begin{figure}[h!]
\centering
\vspace{-0.0in}
%\hspace{-0.1in}
\includegraphics[angle=0,width=5.8in]{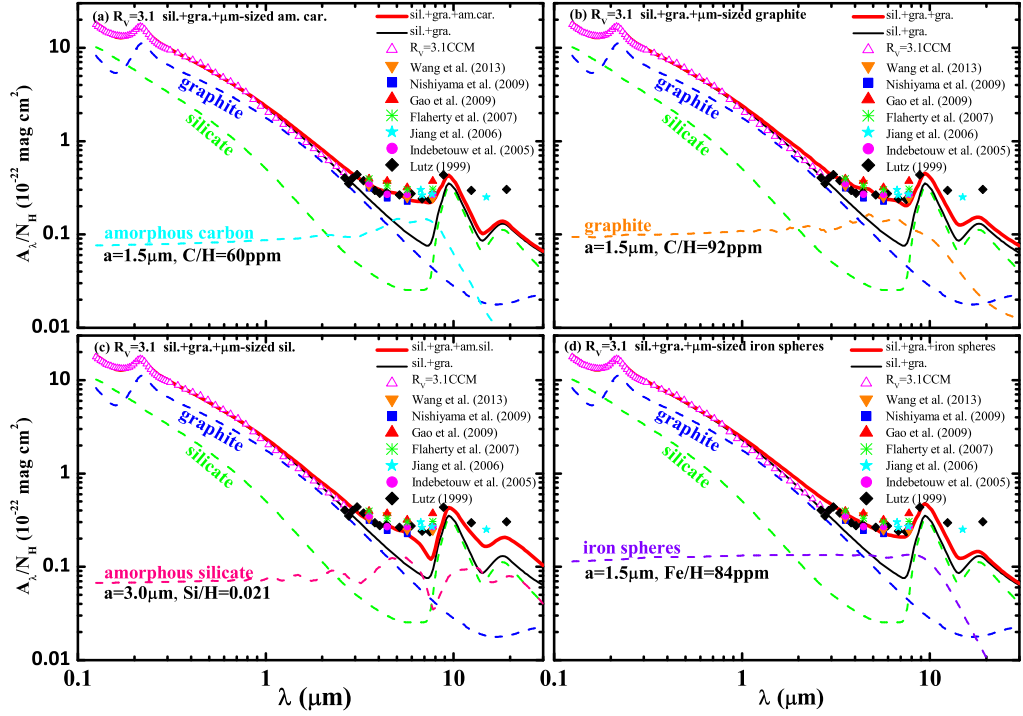}
\vspace{-0.1in}
\caption{\footnotesize
               \label{fig:WLJIR1}
               Fitting the $R_V=3.1$ UV/optical, near- and mid-IR extinction
               with a simple mixture (thin black solid line)
               of amorphous silicate (green dashed line)
               and graphite dust (blue dashed line),
               together with a population of large, $\mu$m-sized dust:
               (a) amorphous carbon of $a\approx 1.5\mum$
                    and C/H\,$\approx$\,60$\ppm$,
               (b) graphite of $a\approx 1.5\mum$
                    and C/H\,$\approx$\,92$\ppm$,
               (c) amorphous silicate of $a\approx 3.0\mum$ and
                    Si/H\,$\approx$\,21,000$\ppm$, and
               (d) iron of $a\approx 1.5\mum$
                   and Fe/H\,$\approx$\,84$\ppm$.
               The thick red solid line plots the model fit,
               while the symbols plot the observed extinction:
               the magenta open triangles plot the $R_V=3.1$
               UV/optical/near-IR extinction, the other symbols
               plot the mid-IR extinction.
               See \S\ref{sec:rst} for details.
               }
\end{figure}

%%%% Figure 5  %%%%%
\begin{figure}[h!]
\centering
\vspace{-0.0in}
%\hspace{-1.0in}
\includegraphics[angle=0,width=5.8in]{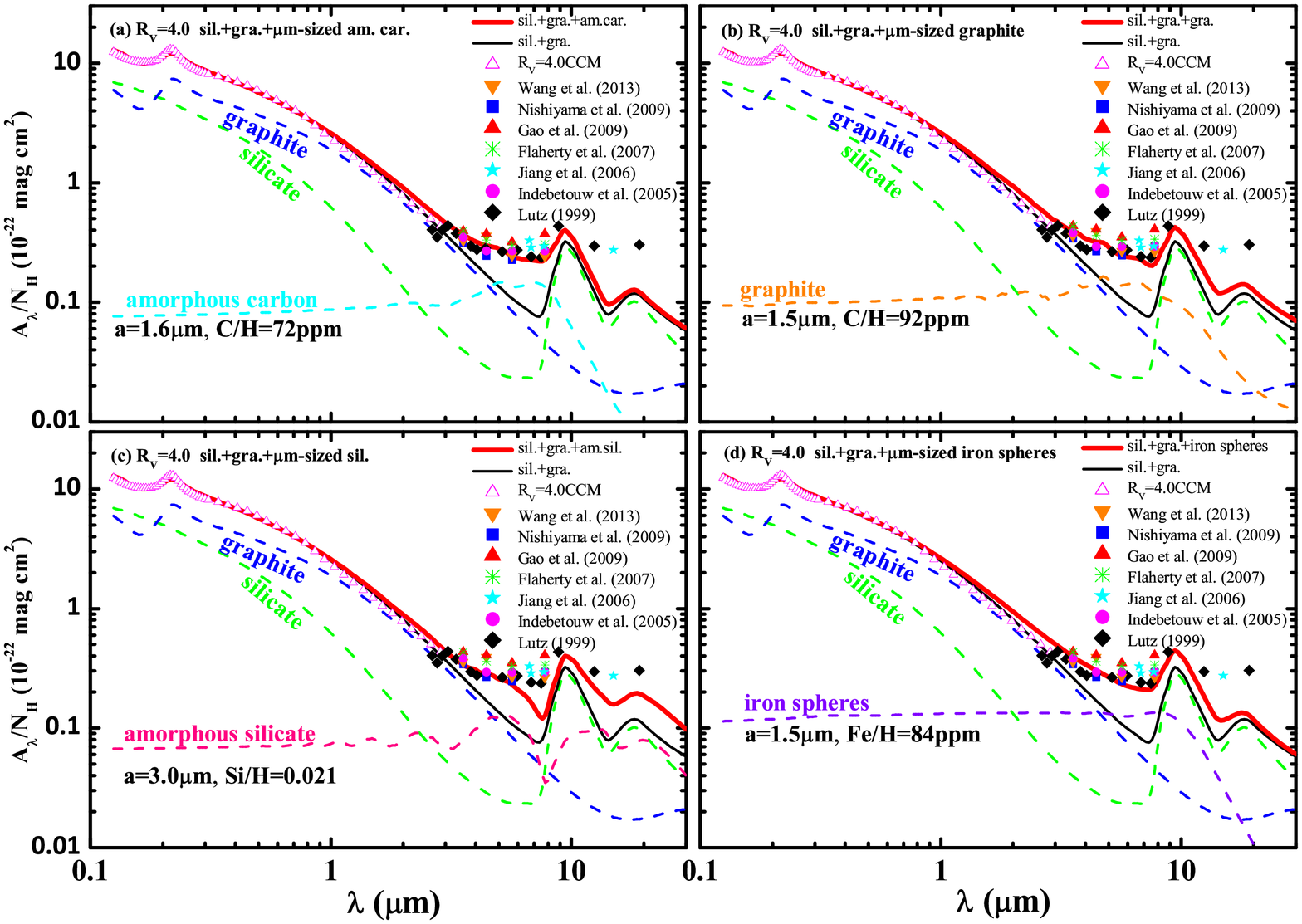}
\vspace{-0.1in}
\caption{\footnotesize
               \label{fig:WLJIR2}
               Same as Figure~\ref{fig:WLJIR1} but for $R_V=4.0$.
               }
\end{figure}

%%%% Figure 6  %%%%%
\begin{figure}[h!]
\centering
\vspace{-0.0in}
%\hspace{-1.8in}
\includegraphics[angle=0,width=5.8in]{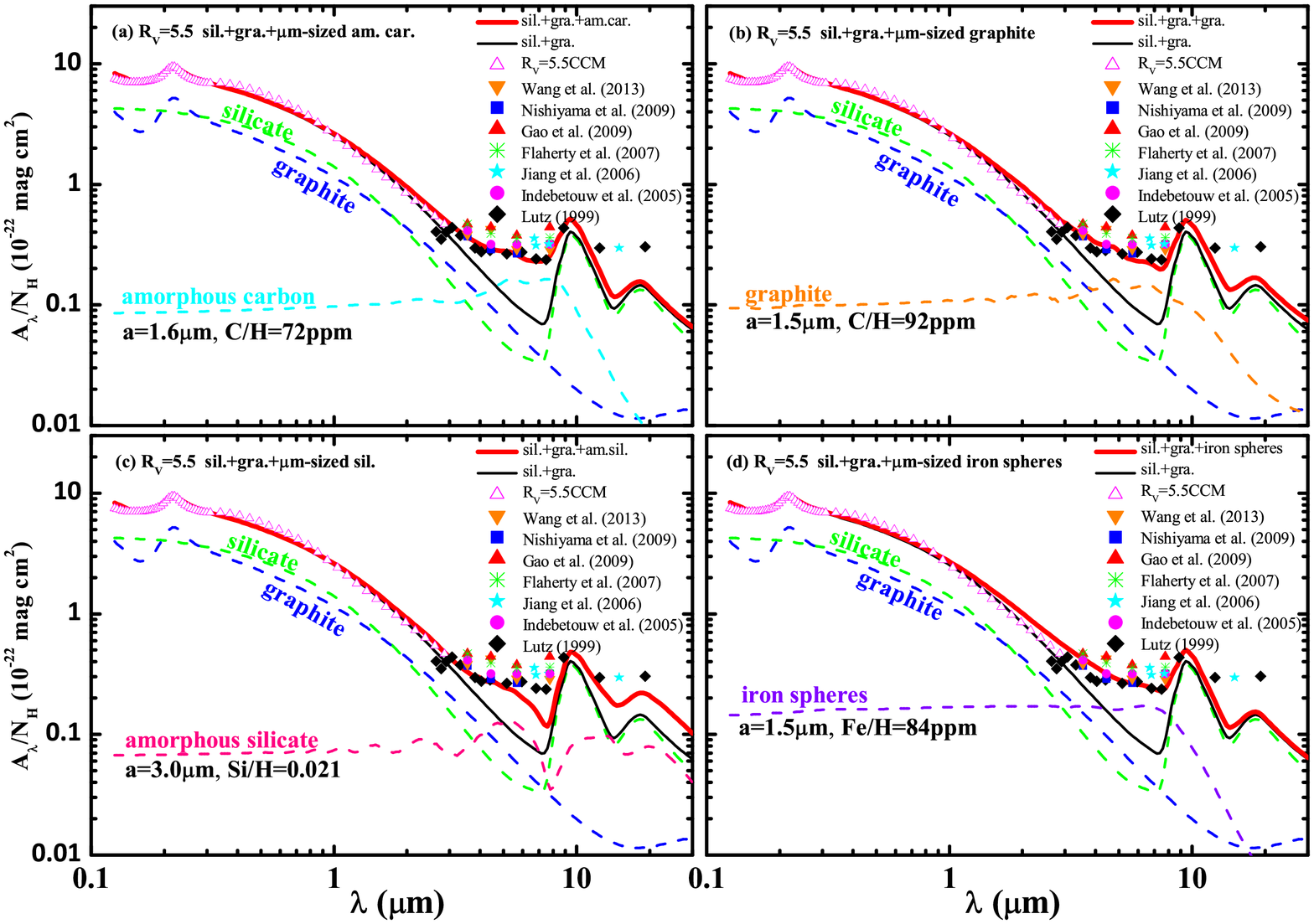}
\vspace{-0.1in}
\caption{\footnotesize
               \label{fig:WLJIR3}
               Same as Figure~\ref{fig:WLJIR1} but for $R_V=5.5$.
               }
\end{figure}

\section{Results and Discussion\label{sec:rst}}
To testify our model, we first fit the UV/optical and near-IR
extinction curve of $R_V=3.1$. We assume that silicate
and graphite have the same size distribution
(i.e., $\alphaS=\alphaC$, $\acS=\acC$).
Figure~\ref{fig:WLJUVOpt} shows the best-fit results.
The best-fit model parameters are listed in Table~\ref{tab:uvpara}.
The abundances of the dust-forming heavy elements
locked up in the dust can be derived
from the following equation:
\begin{eqnarray}
\nonumber
\left[{\rm X}/{\rm H}\right] & = & 4\pi/3 \times
\rho_{\rm X}/m_{\rm H} \times N_{\rm X}/\mu_{\rm X}\\
&& \times \int_{\amin}^{\amax}
   N_{\rm X}^{\prime}\,a^{3-\alphaX}\,\exp\left(-a/\acX\right)\,da ~~,
\end{eqnarray}
where $m_{\rm H} = 1.66 \times 10^{-24}\g$
is the mass of a hydrogen atom,
$\rho_{\rm X}$ and $N_{\rm X}^\prime$
are respectively the mass density
and column density
of the dust species containing element X, and
$N_{\rm X}$ and $\mu_{\rm X}$ are respectively
the number of X atoms in and molecular weight of
a molecule of the dust species containing element X.
We consider elements Si and C and dust species
amorphous silicate and graphite:
$\mu_{\rm C} \approx 12$, $N_{\rm C}=1$,
$\mu_{\rm S} \approx 172$, and $N_{\rm Si}=1$.\footnote{%
   For silicate dust, we assume its chemical composition to
   be MgFeSiO$_4$.
   }
We derive the C and Si abundances
required to be locked up in dust to be
$\cdust\approx 292\ppm$
and $\sidust\approx\mgdust\approx\fedust\approx 34\ppm$
(ppm refers to ``parts per million'').

For the dust-forming element ${\rm X}$,
let $\xism$ be the ``interstellar abundance'' -- the total abundance
of this element in the ISM, both in gas and in dust; and
$\xgas$ be the gas-phase abundance of this element.
The abundance of this element in dust is
$\xdust = \xism-\xgas$.
The gas-phase Mg, Si and Fe abundances are negligible
(i.e. these rock-forming elements are almost completely
depleted from the gas phase; see Li (2005) and references therein).
Therefore, for Si we have $\sidust\approx\siism$.

The interstellar abundance of Si (i.e. $\siism$) is not known.
Traditionally, one often adopts the solar photospheric abundances
(e.g. Asplund et al.\ 2009) of heavy elements
as their interstellar abundances.
However, Lodders (2003) argued that the currently observed solar
photospheric abundances (relative to H) must be lower than
those of the proto-Sun because helium and other heavy elements
have settled toward the Sun's interior since the time of
the Sun's formation some 4.55\,Gyr ago.
She further suggested that protosolar abundances derived
from the photospheric abundances by considering settling effects
are more representative of the solar system elemental abundances.

On the other hand,  it has also been argued that the interstellar
abundances might be better represented by those of B stars and
young F, G stars (because of their young ages)
which are just $\simali$60--70\% of the solar values
(i.e., ``subsolar''; Snow \& Witt 1996, Sofia \& Meyer 2001).
However, Li (2005) showed that if the interstellar abundances are
indeed ``subsolar'' like B stars and young F, G stars,
there might be a lack of raw material to
form the dust to account for the interstellar extinction.
We also note that Przybilla et al.\ (2008) derived
the photospheric abundances of heavy elements for
six unevolved early B-type stars in the solar neighborhood
OB associations and the field using NLTE techniques.
They found that the photospheric abundances of those B stars
are in close agreement with the solar values.
Nieva \& Przybilla (2012) further derived the photospheric
abundances of 29 slowly-rotating early B-type stars.
These stars exhibit $<$10\% abundance fluctuations
and their abundances are also similar to that of the Sun.

The Si abundance of the Sun, proto-Sun,
and early B stars
are respectively
$\sisun\approx 32\ppm$ (Asplund et al.\ 2009),
$41\ppm$ (Lodders 2003),
and $32\ppm$ (Przybilla et al.\ 2008, Nieva \& Przybilla 2012).
The model which best fits the UV/optical extinction requires
$\sidust\approx 34\ppm$, consistent with the solar, proto-Sun
or B stars Si abundance.

For C, it is more complicated since the gas-phase C abundance
is recently under debate. Earlier studies reported
$\cgas\approx 140\ppm$ from the weak intersystem absorption
transition of C\,II] at 2325$\Angstrom$ (Cardelli et al.\ 1996;
Sofia et al.\ 2004). Very recently, Sofia et al.\ (2011)
derived $\cgas\approx 100 \ppm$ for several interstellar
sightlines from the strong transition of C\,II] at 1334$\Angstrom$.
They argued that the oscillator strength for the C\,II] transition
at 2325$\Angstrom$ previously used by Cardelli et al.\ (1996) and
Sofia et al.\ (2004) to obtain $\cgas\approx 140\ppm$
might have been underestimated.
The solar C abundance and proto-Sun C abundance are respectively
$\csun\approx 224\ppm$ (Asplund et al.\ 2009)
and $288\ppm$ (Lodders 2003).
The C abundance of the early B stars
which are thought to be ideal indicators for
the present-day interstellar abundances
since they preserve their pristine abundances
is close to the solar C abundance:
$\cBstar\approx 214\pm20\ppm$ (Przybilla et al.\ 2008)
and $\cBstar\approx 209\pm15\ppm$ (Nieva \& Przybilla 2012).
If the interstellar C abundance is
like that of the early B stars
(i.e., $\cism\approx209\ppm$)
or that of the proto-Sun
(i.e., $\cism\approx288\ppm$),
with the gas-phase C abundance of
$\cgas\approx 100 \ppm$ (Sofia et al.\ 2011) subtracted,
there will be only $\simali$109$\ppm$ or $\simali$188$\ppm$
of C available to make the carbonaceous dust.
However, the model which best fits the UV/optical extinction
requires $\cdust\approx 292\ppm$,
substantially exceeding what would be available to
be locked up in dust.\footnote{%
  We note that, except the composite dust model of
  Mathis (1996) which only requires $\cdust\approx 155\ppm$
  (and $\sidust\approx 31\ppm$, but see Dwek 1997),
  all interstellar grain models consume more C than
  the available value of
  $\simali$109$\ppm$ or $\simali$188$\ppm$ in the ISM:
  $\cdust\approx 194\ppm$
  (and $\sidust\approx 20\ppm$)
  of Li \& Greenberg (1997),
  $\cdust\approx 231\ppm$
  (and $\sidust\approx 48\ppm$) of WD01,
  $\cdust\approx 244\ppm$
  (and $\sidust\approx 36\ppm$)
  of Zubko et al.\ (2004),
  $\cdust\approx 233\ppm$
  (and $\sidust\approx 50\ppm$)
  of Jones et al.\ (2013).
  }

As shown in Figure~\ref{fig:WLJUVOpt}, the silicate-graphite
mixture model with $\sidust\approx 34\ppm$ and
$\cdust\approx 292\ppm$ closely reproduces the
observed UV/optical/near-IR extinction.
However, it fails in fitting the flat 3--8$\mum$ mid-IR extinction
(see  Figure~\ref{fig:WLJIR1}a).
According to light scattering theory, dust absorbs and scatters
starlight most effectively if its size is comparable to the starlight
wavelength. Therefore, the dust which dominates
the mid-IR extinction at $\simali$3--8$\mum$ should be in
micrometer size scale, as inferred from the consideration of
$a\sim \lambda/2\pi$.
This leads us to add an extra population of large,
$\mu$m-sized dust to account for the mid-IR extinction.
We explore the size of the dust
ranging from $a=0.5\mum$ to $a=3.5\mum$.
For simplicity, we only consider
dust of single sizes. In principle, we could
assume a log-normal size distribution
or the WD01-type size distribution
for this $\mu$m-sized dust population.
But we do not expect that a distribution
of dust sizes would affect the conclusion
drawn from the single-size model.
A distribution of sizes would remove the ripple
structures in the extinction curve of
single-sized dust. The observed mid-IR extinction
is commonly derived from broad-band photometry and
therefore could tolerate the wavy ripples.

As shown in Figure~\ref{fig:WLJIR1}a, the $R_V=3.1$ model
together with spherical amorphous carbon dust of radius
of $a\approx1.5\mum$ and C/H\,$\approx$\,60$\ppm$
could closely fit the mid-IR extinction.
These $\mu$m-sized amorphous carbon grains,
with $2\pi a/\lambda\gg 1$,
are ``gray'' in the UV/optical wavelength regime.
Therefore, the addition of $\mu$m-sized dust does not
distort the fit to the observed $R_V=3.1$ extinction curve
provided by the silicate-graphite model.
In total, this model requires $\cdust\approx 352\ppm$
to account for the observed extinction from the UV
to the mid-IR, with $\simali$17\%
of the C atoms locked up in
the $\mu$m-sized amorphous carbon component.

We have also considered $\mu$m-sized graphite,
amorphous silicate, and iron dust.
As shown in Figure~\ref{fig:WLJIR1}b, graphite
of radius of $a=1.5\mum$ is also capable of reproducing
the flat mid-IR extinction. However, the $\mu$m-sized
graphite population requires C/H\,$\approx$\,92$\ppm$.
Therefore, amorphous carbon, with C/H\,$\approx$\,60$\ppm$,
seems more favorable since the $R_V=3.1$ silicate-graphite model
already encounters a ``carbon crisis'' problem:
the model consumes more C/H than what is available.

As shown in Figure~\ref{fig:WLJIR1}c,
amorphous silicate could not fit the mid-IR extinction
for two reasons:
(1) the best-fit model with $a=3\mum$
    predicts a prominent dip
    at $\lambda$$\simali$8$\mum$ which is not seen
    in the observed mid-IR extinction; this dip is due
    to the onset of the 9.7$\mum$ Si--O
    stretch at $\lambda$$\simali$8$\mum$;
(2) this model requires Si/H\,$\approx$\,21,000$\ppm$
    to be locked up in the $\mu$m-sized silicate dust,
    which is far more than the available amount
    of $\siism$\,$\approx$\,32--41$\ppm$ in the ISM.

Figure~\ref{fig:WLJIR1}d shows the fit obtained
with Fe/H\,$\approx$\,84$\ppm$ in iron spheres of
$a=1.5\mum$. Although the fit to the mid-IR extinction
is excellent, this model requires a total depletion of
Fe/H\,$\approx$\,116$\ppm$,\footnote{%
   To account for the UV/optical extinction,
   the submicrometer-sized amorphous silicate
   component consumes Fe/H\,$\approx$\,32$\ppm$.
   }
while the Fe abundance
of the Sun, proto-Sun, and early B stars is only
$\fesun\approx 27.5\ppm$ (Asplund et al.\ 2009),
$\fesun\approx 34.7\ppm$ (Lodders 2003),
and $\feBstar$$\approx$\,28--33$\ppm$
(Przybilla et al.\ 2008, Nieva \& Przybilla 2012),
respectively.

%%%% Below: Table 1 %%%%
\begin{table}[h!]
{\footnotesize
\begin{center}
\caption{\footnotesize
              \label{tab:uvpara}
               Model parameters for fitting the UV/optical and near-IR extinction.
               }
%\vspace{0.2in}
\begin{tabular}{ccccccccc}
\hline \hline
Extinction   & $A_{\Ks}/N_{\rm H}$ & $\alphaS$ & $\alphaC$ & $\acS$ &
$f_{\rm C2S}$ & $\NS$ & $\sidust$ & $\cdust$\\
Type  & ($10^{-23}\magni\cm^2$) & & & ($\mu$m) &
& ($\cm^{-2}$) & (ppm) & (ppm)\\
\hline
$R_V=3.1$  & 6.26 & 3.2 & 3.2 & 0.14 & 0.6 & $4.687\times10^{-24}$ & 34.0 & 292\\
$R_V=4.0$  & 6.85 & 2.7 & 2.7 & 0.12 & 0.6 & $1.669\times10^{-21}$ & 30.7 & 264\\
$R_V=5.5$  & 7.40 & 2.1 & 3.0 & 0.16 & 0.3 & $1.047\times10^{-18}$ & 40.0 & 172 \\
\hline
\end{tabular}
\end{center}
}
\end{table}
%%%% Above: Table 1 %%%%

We have also considered models for the $R_V=4.0$ and $R_V=5.5$
extinction curves since the flat mid-IR extinction has also been seen
in dense regions (see \S\ref{sec:status}).
For the $R_V=4.0$ case, we also assume that both silicate and graphite
have the same size distribution. The results are shown in
Figure~\ref{fig:WLJIR2} and Table~\ref{tab:uvpara}.
It is seen that both amorphous carbon of $a\approx1.6\mum$
and graphite of $a\approx1.5\mum$ could fit the flat mid-IR
extinction. Again, amorphous carbon is preferred since it requires
C/H\,$\approx$\,72$\ppm$, less than that of graphite of
C/H\,$\approx$\,92$\ppm$.
For the $R_V=5.5$ case, we could not fit the UV/optical/near-IR
extinction if we assume the same size distribution for both silicate
and graphite. We therefore set $\acS=\acC$, but allow silicate and
graphite to have different $\alpha$ values (i.e., $\alphaS\ne\alphaC$).
The best-fit results are shown in Figure~\ref{fig:WLJIR3} and the
model parameters are tabulated in Table~\ref{tab:uvpara}.
Similar to the $R_V=3.1$ and $R_V=4.0$ models,
$\mu$m-sized amorphous carbon and graphite could closely
fit the mid-IR extinction, with amorphous carbon being preferred
since it does not consume as much C/H as graphite.

Finally, we note that it is not clear how
$\mu$m-sized interstellar dust is formed.
However, there are several pieces of evidence
suggesting its presence in the ISM:
(1) measurements by dust impact detectors
    on the interplanetary spacecrafts
    {\it Ulysses} and {\it Galileo}
    appear to indicate a substantial flux of
    interstellar particles with
    masses $>$\,10$^{-12}\g$
    (corresponding to $a>0.4\mum$ for silicate
     and $a>0.5\mum$ for graphite)
     entering the heliosphere
     (see Landgraf et al.\ 2000,
          Kr\"uger et al.\ 2007);
(2) Taylor et al.\ (1996) reported radar detection of
$a\approx 30\mum$ particles
entering the Earth's atmosphere
on solar-hyperbolic trajectories
implying that they are arriving from interstellar space.
Socrates \& Draine (2009) discussed the detectability
of very large interstellar grains of $a$$\simali$1\,mm
(``pebble'') through optical scattered light halos.

\section{Summary}
The mid-IR extinction curves of a wide variety of interstellar regions
(including both diffuse and dense environments)
exhibit a flat or ``gray'' behavior in the wavelength region of
$\simali$3--8$\mum$.
This flat mid-IR extinction is hardly accounted for
by the standard MRN silicate-graphite dust model.
To explain the flat mid-IR extinction, we have considered
various dust sizes and species,
including amorphous silicate, graphite,
amorphous carbon, and iron spheres.
we find that $\mu$m-sized amorphous carbon dust
best fits the flat mid-IR extinction.
The observed extinction from the UV to
the mid-IR could be closely reproduced
by a mixture of
submicrometer-sized amorphous silicate dust,
submicrometer-sized graphitic dust,
and micrometer-sized amorphous carbon dust.
However, this mixture requires a solid-phase
C abundance of C/H\,$\approx$\,352\,ppm,
considerably exceeding what could be available
in the ISM.

\section*{Acknowledgements}
\label{acknowledgements}
We thank J.~Gao, A.~Mishra,
and the anonymous referees
for helpful comments/suggestions.
We are supported in part by NSFC\,11373015 and 11173007,
NSF AST-1109039, NASA NNX13AE63G,
and the University of Missouri Research Board.

%%\section*{References}

\end{document}